\documentclass[aps,prl,superscriptaddress,reprint]{revtex4-2}
\usepackage{amsmath}
\usepackage{amssymb}
\usepackage{graphicx}
\usepackage[colorlinks,urlcolor=blue]{hyperref}
\usepackage{url}
\usepackage{hyperref}
\usepackage{color,xcolor}
\usepackage{epstopdf}
\usepackage{float}
\usepackage{ulem}
\usepackage[percent]{overpic}
\usepackage{siunitx}

\begin{document}

\title{Altermagnetism in Folded Iron Chain Systems}
\author{Ling-Fang Lin}
\affiliation{School of Physics, Southeast University, Nanjing 211189, China}
\affiliation{Department of Physics and Astronomy, University of Tennessee, Knoxville, Tennessee 37996, USA}
\author{Yang Zhang}
\email{yzhang@utk.edu}
\affiliation{Department of Physics and Astronomy, University of Tennessee, Knoxville, Tennessee 37996, USA}
\author{Satoshi Okamoto}
\affiliation{Materials Science and Technology Division, Oak Ridge National Laboratory, Oak Ridge, Tennessee 37831, USA}
\author{Adriana Moreo}
\affiliation{Department of Physics and Astronomy, University of Tennessee, Knoxville, Tennessee 37996, USA}
\affiliation{Materials Science and Technology Division, Oak Ridge National Laboratory, Oak Ridge, Tennessee 37831, USA}
\author{Lun-Hui Hu}
\email{lunhui@zju.edu.cn}
\affiliation{Center for Correlated Matter and School of Physics, Zhejiang University, Hangzhou 310058, China}
\author{Elbio Dagotto}
\affiliation{Department of Physics and Astronomy, University of Tennessee, Knoxville, Tennessee 37996, USA}
\affiliation{Materials Science and Technology Division, Oak Ridge National Laboratory, Oak Ridge, Tennessee 37831, USA}

\date{\today}

\begin{abstract}
Altermagnetism—moment-compensated magnetism with momentum-dependent spin splitting—has so far been realized primarily in N\'{e}el-type antiferromagnets. Here, we demonstrate altermagnetism in folded iron-chain systems with block-type magnetic order (e.g.,$\uparrow \uparrow \downarrow \downarrow$). In the strictly one-dimensional (1D) case, altermagnetism is forbidden by the $[C_2||{\cal T}]$ symmetry, but it can emerge when such chains are stacked into a quasi-1D array. To generate altermagnetic spin splitting via this chain-stacking mechanism, two essential ingredients are required: a block-type spin order that spontaneously breaks the combined parity-time reversal symmetry, and finite inter-chain hopping. Using first-principles calculations, we identify the K$_3$Fe$_2$Se$_4$ family as a promising platform that realizes this mechanism, hosting a $d$-wave altermagnet even without lattice $C_{4}$ symmetry. Moreover, owing to its lower symmetry, the block-type folded chain exhibits an isotropic altermagnetic spin splitting, which may manifest as an altermagnetic wrapping effect. Our findings unify real-space block magnetism and reciprocal-space altermagnetism, establishing folded iron chains as a tunable platform for zero-moment spintronic applications.
\end{abstract}

\maketitle

\noindent{\it \color{blue} Introduction.--} Altermagnetism is an emerging quantum magnetic phenomenon that lies beyond the conventional dichotomy of ferromagnetism and antiferromagnetism, and has rapidly gained attention due to its rich fundamental physics and potential spintronic applications \cite{Smejkal:prx1,Smejkal:prx2,Mazin:prx,Song:nrm,Naka:nc,Smejkal:sa,Ahn:prb,Yuan:prb,Hayami:prb,Ma:nc,Mazin:pnas,Gao:nsr2025,Liu:prx,Antonenko:prl2025,Fernandes:prb2024,Zhou:prb2025,Li:pnas2024,Jungwirth:arXiv2024,Fender:jacs2025,Naka:prb2021,Shao:NatComm,Zhang:prl2024Predictable,Bhowal:prx,Leeb:prl2024,Duan:prl2025,Gu:prl2025,Zhu:NanoLett2025,Maier:prb2023Weak,Bose:prb2024,Kaushal:arXiv}. In contrast to ferromagnets, which exhibit a net magnetization and uniform Zeeman spin splitting across the Brillouin zone (BZ), and conventional antiferromagnets, which maintain Kramers degeneracy due to compensated collinear magnetic order, altermagnets feature compensated magnetic moments with non-relativistic spin splitting across the BZ. This unique band splitting, proposed in early phenomenological studies to occur even in the absence of spin-orbit coupling~\cite {Wu:PRL2004,Wu:PRB2007}, can be realized in magnetic materials, where it stems from the magnetic symmetry of the crystal structure \cite{Jungwirth:nature}.

A defining characteristic of altermagnets is that their opposite spin sublattices are related by crystal symmetries such as rotation, mirror, glide, or screw operations—rather than inversion or simple translation. This symmetry-induced mechanism results in alternating spin splitting throughout distinct regions of the BZ \cite{Smejkal:prx1,Liu:prx,Xiao:prx,Chen:prx,Jiang:prx}, giving rise to unconventional transport phenomena. For example, altermagnets can exhibit large anomalous Hall and Nernst effects—even in the absence of net magnetization or spin-orbit coupling—challenging traditional assumptions in spintronics \cite{Smejkal:sa,Feng:ner,Han:sciadv2024b,Han:arxiv2024a}. A growing list of candidate altermagnetic materials has been proposed theoretically \cite{Gao:nsr2025,Smejkal:prx1}, and a number of them have recently been confirmed experimentally \cite{Krempasky:nat2024,Fedchenko:sciadv2024,Lee:prl2024,Ding:prl2024,Zeng:adv2024,Yang:natcomm2025,Reimers:natcomm2024,Osumi:prb2024,Liu:prl2024chiral,Liu:prl2024,Jiang:nphys2025,Zhang:arxiv2024,Feng:ner,Han:sciadv2024b,Han:arxiv2024a}. Most of these systems host conventional N\'{e}el collinear antiferromagnetic order ($\uparrow \downarrow \uparrow \downarrow$), in which each spin is strictly antiparallel to its nearest neighbor.

As a counterpart of altermagnetism in real space, there is an intriguing real-space magnetic texture known as the block magnetic state (e.g., $\uparrow \uparrow \downarrow \downarrow$), which is commonly observed in iron-based superconducting materials such as BaFe$_2$Se$_3$ and $A_2$Fe$_4$Se$_5$ ($A$ = K, Rb, Cs, Tl) \cite{Mourigal:prl2015,Ye:prl2011}. In this block state, magnetic moments align ferromagnetically within small blocks, but adjacent blocks are coupled antiferromagnetically. The origin of this block-type order has been attributed to the competition between nearest-neighbor ferromagnetic and next-nearest-neighbor antiferromagnetic exchange interactions, further enriched by lattice distortions, orbital-selective correlations, and strong electronic anisotropy \cite{Dagotto:rmp2013, Lin:prl2021, Lin:prb2022, Lin:cp2023}.

Given the rise of altermagnetism and its profound connection to crystal symmetry and spin texture, a compelling question arises: Can altermagnetic behavior manifest in block-type magnetic systems? If so, what are the microscopic mechanisms and symmetry requirements for realizing altermagnetism in these spatially modulated magnetic configurations? What novel electronic or transport phenomena might such a state exhibit?

\begin{figure*}[t]
\centering
\includegraphics[width=0.9\textwidth]{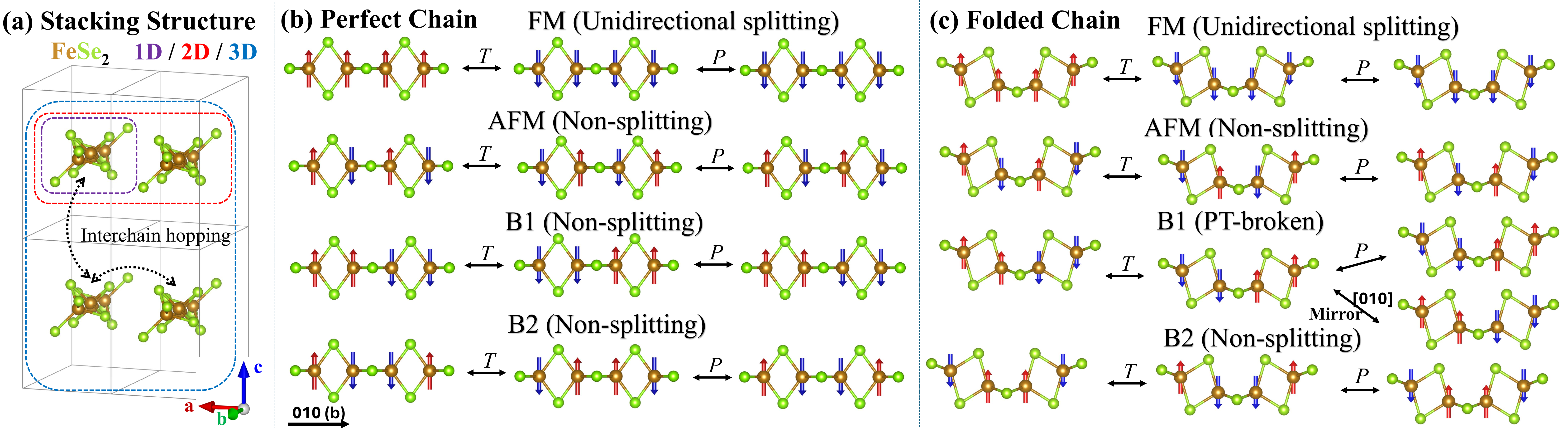}
\caption{ (a) Schematic illustration of the FeSe$_2$ stacking structures visualized via the VESTA code~\cite{Momma:vesta}, showing the evolution from isolated 1D chains to stacked two-dimensional (2D) and three-dimensional (3D) lattices through interchain hopping. Schematics of time-reversal ($T$) and inversion/translation ($P$) symmetry operations applied to different magnetic configurations in (b) perfect and (c) folded iron chains with interchain hopping. Four representative magnetic states are analyzed: ferromagnetic (FM), antiferromagnetic (AFM), block type 1 (B1), and block type 2 (B2). Brown and green spheres represent Fe and Se atoms, respectively, with Se forming FeSe$_4$ tetrahedra. Note that altermagnetic splitting emerges only with finite inter-chain hopping.
}
\label{PTsymmetry}
\end{figure*}

In this Letter, we address these fundamental questions by proposing a symmetry-guided framework to realize altermagnetism in block magnetic states. We begin by identifying the necessary magnetic and crystallographic symmetry conditions that support non-relativistic spin splitting in the absence of net magnetization. Later, we employ first-principles calculations to examine the evolution of the electronic structure under operations such as chain rotation and decreasing interchain coupling. Then, we explore possible
real candidate materials where block-state altermagnetism may be observed, providing experimental guidance for future validation. Finally, we also discuss the interesting transport phenomena of the block-magnetic altermagnet. Our results pave the way toward unifying real-space magnetic block states with reciprocal-space altermagnetic signatures and open a new frontier in symmetry-driven quantum magnetism.

\noindent {\it \color{blue} The General Mechanism.--} Motivated by the stacking-construction paradigm well-established for weak topological insulators~\cite{Fu:prl2007}, here we realize altermagnetism in a block magnetic state by introducing controlled inter-chain coupling to a minimal one-dimensional (1D) folded FeSe$_4$ chain [see Fig.~\ref{PTsymmetry}(a)]. This provides a general symmetry-guided route to engineering altermagnetic phases in quasi-1D systems.

In the absence of spin-orbit coupling, a collinear antiferromagnetic order must respect the $[C_2||{\cal T}]$ symmetry~\cite{Zhuang:arxiv2025}, where $C_2$ is a spin-flip operator acting solely on the spin subspace and ${\cal T}$ is the spinful time-reversal operator. Consequently, altermagnetism in any dimension requires even-parity band dispersions $E_{\sigma}(\vec{k})=E_{\sigma}(-\vec{k})$ with $\sigma=\uparrow$ and $\downarrow$. In the strict 1D case, altermagnetism is forbidden: the $[C_2||M_x]$ symmetry enforces zero net magnetization and constrains the bands to satisfy $E_{\uparrow}(k_x)=E_{\downarrow}(-k_x)$, which together with the even-parity condition yields fully spin-degenerate bands. In 2D, introducing the inter-chain hopping leads to $E_{\uparrow}(k_x,k_y)=E_{\downarrow}(-k_x,k_y)$. Combined with $E_{\sigma}(k_x,k_y)=E_{\sigma}(-k_x,-k_y)$, this gives $E_{\uparrow}(k_x,k_y)=E_{\downarrow}(k_x,-k_y)$, yielding symmetry-enforced altermagnetic nodal lines along $k_x=0$ and $k_y=0$—the hallmark of $d$-wave altermagnetism even without lattice $C_4$ symmetry. The 3D case follows naturally by stacking 2D layers, producing analogous nodal lines in the $k_x$-$k_{yz}$ plane ($k_y=k_z$); see details in the Supplementary Materials (SM)~\cite{supp}.

We now examine the FeSe$_2$ chain composed of edge-sharing FeSe$_4$ tetrahedra, the basic building block of the chain-stacking construction. As illustrated in Figs.~\ref{PTsymmetry}(b-c), we examine four representative magnetic configurations—ferromagnetic (FM), antiferromagnetic (AFM), and two distinct block-type patterns (B1 and B2)—within both ideal (straight) and folded geometries. The folded structure introduces a substantial bending angle between every two FeSe$_4$ units, fundamentally altering the symmetry landscape of the system.

In the FM configuration, both ideal (unfolded)  and folded chains break ${\cal T}$, resulting in Kramers degeneracy lifting and a uniform spin splitting across the whole BZ. In contrast, the AFM and B2 states preserve Kramers degeneracy through a combination of time-reversal symmetry and spatial symmetries such as inversion and/or lattice translation, regardless of whether the structure is ideal or folded.

In the ideal chain, the B1 and B2 magnetic patterns are symmetry-equivalent, both preserving ${\cal P} {\cal T}$ symmetry and leading to spin-degenerate band structures. However, this equivalence is broken in the folded geometry. Notably, the B1 configuration in the folded chain breaks ${\cal P} {\cal T}$ due to the structural distortion introduced by the bending. While ${\cal P}$ and mirror symmetry are preserved, the absence of ${\cal P} {\cal T}$ symmetry gives rise to altermagnetic spin splitting in momentum space once inter-chain hopping is turned on. Thus, altermagnetism in the iron chains requires two key ingredients: {\it a block-spin texture} that breaks ${\cal P} {\cal T}$ symmetry, and {\it a finite inter-chain hopping}. This highlights the delicate interplay between lattice geometry and magnetic symmetry, revealing how subtle structural modulations can enable altermagnetism.

\begin{figure}[t]
\centering
\includegraphics[width=0.44\textwidth]{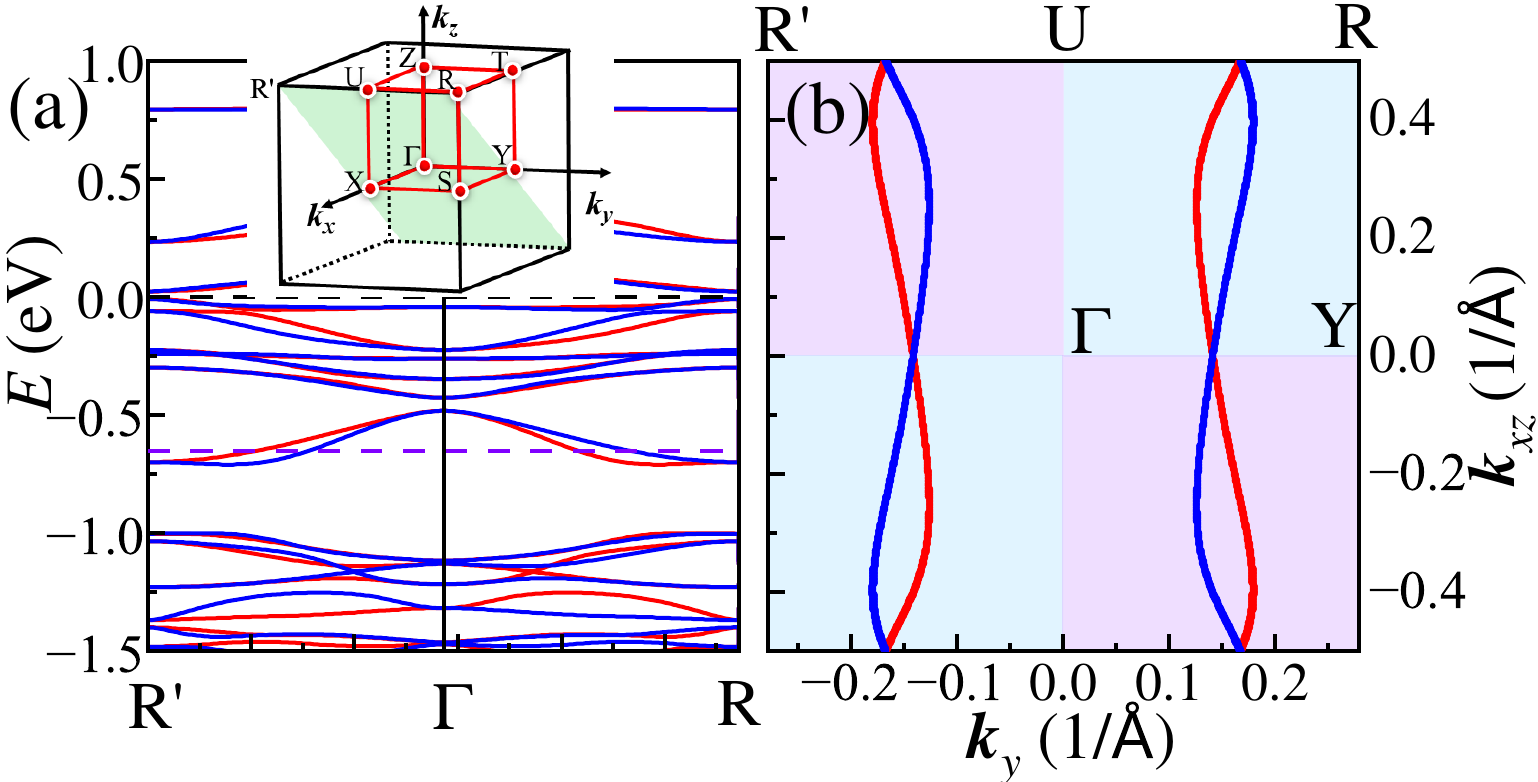}
\caption{(a) Spin-resolved band structure of the B1 magnetic configuration in a folded iron chain, plotted along the high-symmetry path R'–$\Gamma$–R. (b) Fermi surface corresponding to a lower chemical potential $\mu = -0.65$ eV, indicated by the dashed purple line in (a).}
\label{band_fs}
\end{figure}

To further illustrate this, we consider a simplified 3D crystal model with inter-chain couplings [see Fig.~\ref{PTsymmetry}(a)]. Here, the lattice constants (in units of \AA) are $a=7.400$ and $c=11.937$ for both folded and ideal structures, while $b=11.296$ and $11.546$ for the folded and ideal structures, respectively. We present the calculated spin-resolved band structure~\cite{Kresse:prb1996,Kresse:cms1996,Perdew:prl1996,Dudarev:prb1998,Kresse:prb1993,Blochl:Prb} and Fermi surface (chemical potential $\mu$=-0.65 eV) for the folded B1 configuration as shown in Fig.~\ref{band_fs}. A clear alternating spin splitting is observed along the high-symmetry path R'–$\Gamma$–R, confirming the emergence of non-relativistic spin polarization despite the absence of net magnetization. Owing to the 2/m point group symmetry of the folded chain crystal, the spin splitting exhibits a $d$-wave character in momentum space in agreement with the general symmetry analysis discussed above. This is further confirmed by the spin-resolved Fermi surfaces shown in Fig.~\ref{band_fs}(b).

\begin{figure}[t]
\centering
\includegraphics[width=0.42\textwidth]{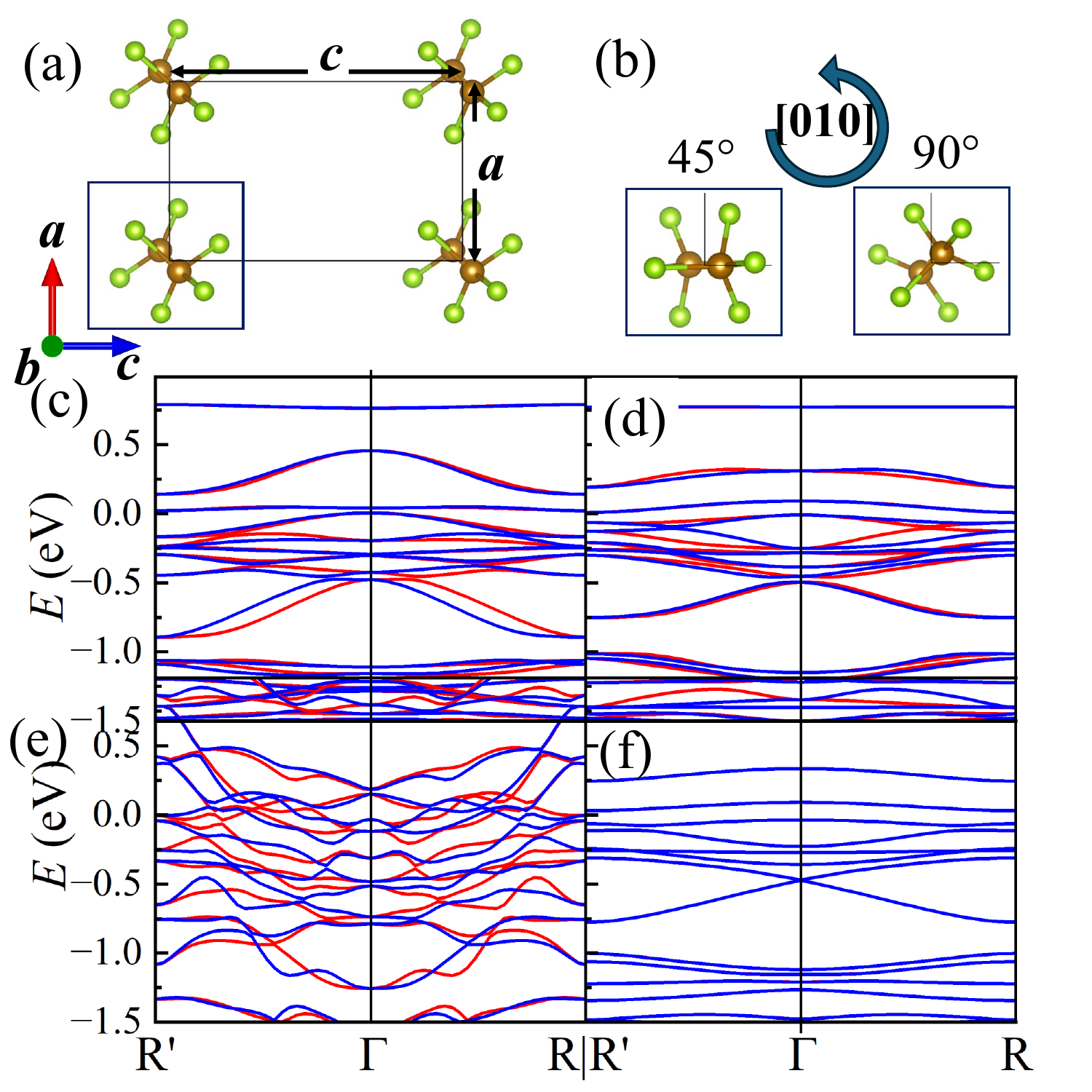}
\caption{(a) Crystal structure of the folded single-chain system with the chain direction aligned along the $b$ axis. The lattice constants along the $a$ and $c$ directions are denoted as $a$ and $c$, respectively. (b) Schematic illustration of rotating the chain around the $b$ axis by $45^\circ$ and $90^\circ$, while keeping the lattice box fixed. (c, d) Corresponding spin-resolved band structures for the rotated chain configurations: (c) $45^\circ$ and (d) $90^\circ$. (e, f) Band structures for two different lattice constants: (e) $a = c = 5.000$ \AA ~  and (f) $a = c = 15.000$ \AA. Note that the interchain hopping is effectively enhanced in panel (e) and completely suppressed in panel (f).}
\label{interchain}
\end{figure}

\noindent {\it \color{blue} Interchain Interaction Effect.--} To further explore the tunability and robustness of altermagnetism, we investigate its evolution under variations in interchain orientation and spacing, as illustrated in Fig.~\ref{interchain}. The starting point is the primitive folded chain configuration [Fig.~\ref{interchain}(a)], whose electronic structure has been previously shown in Fig.~\ref{band_fs}. We first consider the effect of rotating the chain about the $b$-axis by $45^\circ$ and $90^\circ$, while keeping the lattice dimensions fixed [Fig.~\ref{interchain}(b)]. The corresponding spin-resolved band structures [Figs.~\ref{interchain}(c, d)] reveal that the hallmark alternating spin splitting along the R'–$\Gamma$–R path persists, indicating that the altermagnetic behavior is robust against such rotations. Interestingly, the magnitude of spin splitting varies non-monotonically with the rotation angle: at a fixed chemical potential $\mu = -0.65$ eV, the splitting increases when the rotation reaches $45^\circ$, then diminishes upon further rotation to $90^\circ$.

We next examine the role of interchain coupling by varying the lattice constants $a$ and $c$ while preserving the folded chain geometry. When the interchain distance is reduced to $a = c = 5.000$~\AA ~with increasing interchain hopping, the band structure becomes more dispersive, and the alternating spin splitting is enhanced across multiple bands [Fig.~\ref{interchain}(e)]. In contrast, when the interchain distance is increased to $a = c = 15.000$~\AA ~with fully suppressed interchain hopping, effectively isolating the chains, the spin splitting vanishes entirely [Fig.~\ref{interchain}(f)]. This behavior underscores the critical role of interchain interaction and crystal symmetry in stabilizing the altermagnetic splitting in the block magnetic phase, supporting the discussion of the general chain-stacking mechanism [see End Matter]. We also present effective tight-binding model calculations analyzing the role of inter-chain hopping in altermagnetism in the SM~\cite{supp}.

\begin{figure}
\centering
\includegraphics[width=0.44\textwidth]{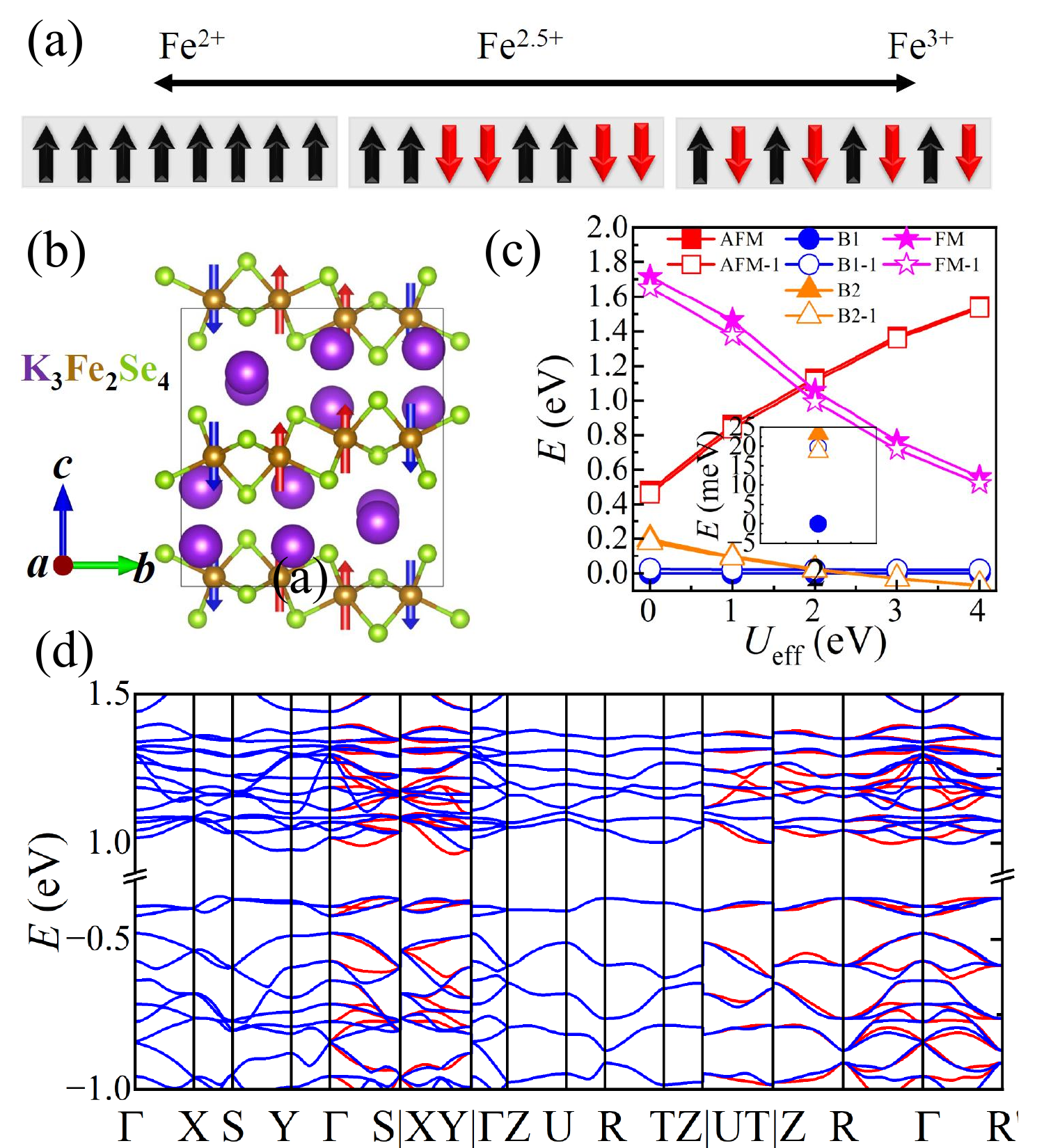}
\caption{(a) Schematic phase diagram illustrating typical magnetic configurations in iron-based chains as a function of electron occupation. (b) B1-type magnetic configuration of K$_3$Fe$_2$Se$_4$; (c) Total energy comparison of various collinear magnetic patterns obtained from first-principles calculations, and here the energies of B1-type are set as reference. Note: only collinear magnetic configurations are considered. Since each unit cell contains two chains, the “*-1” notation indicates antiferromagnetic coupling between the chains, while the absence of “-1” implies ferromagnetic coupling. Inset: a more detailed energy comparison between the B1 and B2 patterns at $U$ = 2 eV. (d) Electronic band structure corresponding to the B1 magnetic configuration, showing features consistent with altermagnetism.}
\label{example}
\end{figure}

\noindent {\it \color{blue} Real Example Materials.--} Then, how to realize altermagnetism in real materials based on the block state idea? In iron chain systems, three typical magnetic configurations are associated with different electron occupations of Fe ions, corresponding to Fe$^{2+}$ ($3d^6$), Fe$^{2.5+}$ ($3d^{5.5}$), and Fe$^{3+}$ ($3d^5$), respectively [see Fig.~\ref{example}(a)]. Representative materials for these cases include Ce$_2$O$_2$FeSe$_2$ (Fe$^{2+}$), K$_3$Fe$_2$Se$_4$ (Fe$^{2.5+}$), and BaFe$_2$Se$_4$ (Fe$^{3+}$). In Ce$_2$O$_2$FeSe$_2$, the ``half-full'' mechanism stabilizes FM order along the chain direction \cite{McCabe:prb,Lin:prl2021}. For BaFe$_2$Se$_4$, the strong Anderson superexchange mechanism leads to a robust AFM ground state \cite{Zhang:prb}. The intermediate case of Fe$^{2.5+}$ is of particular interest, as both the FM ``half-full'' mechanism and AFM superexchange are relevant. Their competition gives rise to a characteristic block magnetic configuration, specifically the $\uparrow$-$\uparrow$-$\downarrow$-$\downarrow$ (B1-type) pattern as predicted in Ref.~\cite{Lin:prb2022}.

Recent neutron scattering experiments have confirmed the collinear B1-type magnetic configuration along the chain direction in the material K$_3$Fe$_2$Se$_4$ [see Fig. \ref{example} (b)] ~\cite{Gao:prl}. Even if the folding introduces perturbations that locally break spatial inversion—such as the staggered altermagnetic exchange field discussed in the SM~\cite{supp}—the magnetic moments remain strictly collinear. Therefore, we consider only collinear magnetic configurations. The system contains two magnetic chains per unit cell, each exhibiting collinear spin alignment. The angle between the chains is approximately $90^\circ$, which—according to Fig.\ref{interchain}—can modulate the band structure while preserving the key features of altermagnetism.

The total energy comparisons are shown in Fig.\ref{example}(c). Among various magnetic patterns, the energy difference between ferromagnetically and antiferromagnetically coupled chains is relatively small, suggesting weak interchain exchange interactions. Within each chain, the FM and AFM states have higher energies, while B1 and B2 patterns are energetically competitive. The ground states vary depending on the $U$ value. Specifically, at $U = 2$ eV, the B1 configuration is energetically favored over B2, making it the ground state. The corresponding band structure of the B1 configuration is presented in Fig.\ref{example}(d), displaying the features of an altermagnet.

These interesting block-type magnetic states may be closely related to the physics of orbital-selective Mott phases, as discussed in our previous work on the Fe$^{2.5+}$ chain~\cite{Lin:prb2022}. Another notable feature is the nonsymmorphic symmetry. In the unit cell, the two chains are related by two glide symmetries. As shown in Fig.~\ref{example}(d), the spin-split bands are twofold degenerate along the Z–R line for the B1 spin configuration. This degeneracy arises from the combination of the glide symmetry—which combines a mirror reflection about the [001] axis with a fractional translation—and the combined inversion and $[C_2||T]$. The degeneracy is lifted when $k_z$ deviates from $\pi$, such as along the $\Gamma$-S line.

Recently, another work also utilized block magnetic order to break the combined ${\cal P}{\cal T}$ symmetry and realize altermagnetism in $A_2$Fe$_4$Se$_5$ under pressure~\cite{Li:prb25}. The underlying physical insight is based on the 2D Lieb lattice~\cite{Kaushal:arXiv,Zhu:prl25}, which enables non-relativistic momentum-dependent spin splitting. However, our work establishes a general 1D chain-construction framework for engineering 2D or 3D altermagnetism, where the inter-chain coupling and block order together stabilize altermagnetic splitting. By systematically stacking folded 1D chains with tunable inter-chain coupling, our approach offers a transferable design principle that can be generalized to a broad family of quasi-1D materials.

Besides K$_3$Fe$_2$Se$_4$, the family of $A_3$Fe$_2X_4$ compounds ($A$ = Na, K, Rb, Cs, Tl; $X$ = S, Se) consists of serial iron-chain materials with folded chains, providing a rich platform to explore altermagnetism in folded-chain systems \cite{Stueble:zaac2017,Harrison:ccr,Bronger:jalloy,Klepp:kryst2000,Bronger:zaac1999}. To date, only a few members of this family have been investigated in detail regarding their magnetic properties \cite{Ensling:hfi,Welz:jpcs1989,Ghazanfari:rscadv2022}. Based on the limited description available for Tl$_3$Fe$_2$S$_4$ \cite{Welz:jpcs1989}, this compound exhibits a collinear block magnetic state, though it remains unclear whether it corresponds to the B1 or B2 type. Na$_3$Fe$_2$Se$_4$ and Na$_3$Fe$_2$S$_4$ exhibit one-dimensional antiferromagnetic order within the chains, while the detailed spin configuration remains undetermined~\cite{Ensling:hfi}. In this case, collinear block states, noncollinear physics, or canted spins with small net moments may be realized in some 324-type materials, which warrant further investigation.

\begin{table*}[t]
\centering
\caption{ Comparison of symmetry and transport properties among representative $d$-wave altermagnets (AM).}
\begin{tabular}{c|c|c|c|c|c}
\hline\hline
$d$-wave AM &
Symmetry &
Spin-splitting &
Charge conductivity &
Spin conductivity &
Spin polarization \\
\hline

RuO$_2$ films 
&
\begin{tabular}{c}
$[C_2||C_{4z}]$ \\
$[C_2||M_{x(y)}]$ \\
$[E||M_{xy(\bar{x}y)}]$
\end{tabular}
&
$J_0 k_xk_y + J_1(k_xk_y^3+k_x^3k_y)$
&
$\sigma_{xx}=\sigma_{yy}$
&
$\sigma^{\rm spin}_{xy}
=\sigma^{\rm spin}_{yx}$
&
$\chi^{s_z}_{xy} = \chi^{s_z}_{yx}$
\\[6pt]
\hline
\begin{tabular}{c}
Monolayer \\
KV$_2$Se$_2$O family
\end{tabular}
&
\begin{tabular}{c}
$[C_2||C_{4z}]$ \\
$[C_2||M_{xy(\bar{x}y)}]$ \\
$[E||M_{x(y)}]$
\end{tabular}
&
$J_0(k_x^2-k_y^2) + J_1(k_x^4-k_y^4)$
&
$\sigma_{xx}=\sigma_{yy}$
&
$\sigma^{\rm spin}_{xx}
=-\sigma^{\rm spin}_{yy}$
&
$\chi^{s_z}_{xx}=-\chi^{s_z}_{yy}$
\\[6pt]
\hline
Folded chain
&
$[C_2||M_x]$
&
$J_0k_xk_y + J_1 k_xk_y^3 + J_2 k_x^3k_y$
&
$\sigma_{xx}\neq\sigma_{yy}$
&
$\sigma^{\rm spin}_{xy}
=\sigma^{\rm spin}_{yx}$
&
$\chi^{s_z}_{xy}=\chi^{s_z}_{yx}$
\\

\hline\hline
\end{tabular}
\label{Table1}
\end{table*}

\noindent {\it \color{blue} Discussion and Conclusion.--} 
We briefly discuss the transport properties of this block-magnetic altermagnet based on tight-binding model calculations (see the SM~\cite{supp}). The chain-stacking constructed altermagnet exhibits a rich set of symmetry-selected responses, distinct from those of both ferromagnets and conventional antiferromagnets. We further compare our folded-chain system with other representative $d$-wave altermagnets, including RuO$_2$ thin films~\cite{Smejkal:sa,Feng:ner,Berlijn:prl,Zhu:prl22,Hiraishi:prl24} and the monolayer KV$_2$Se$_2$O family~\cite{Jiang:nphys2025,Zhang:arxiv2024,Wang:arxiv2025-V,Fu:arxiv2025-V,Yang:arxiv2025-V,Mu:arxiv2025-V}. Unlike these systems, which possess lattice $C_4$, our chain-stacking altermagnet naturally lacks $C_4$ symmetry and thus has a lower-symmetry crystalline environment. Note that K$_3$Fe$_2$Se$_4$ realizes bulk altermagnetism without layer-parity constraints, in contrast to the hidden altermagnetism in the bulk KV$_2$Se$_2$O families~\cite{Sun：prb2026,Yang:arxiv2025,Li:arxiv2025}. 
Nevertheless, our general chain-stacking mechanism can be straightforwardly extended to construct hidden altermagnetism~\cite{Guo:fp,Matsuda:prl,Meier:prl}.
As shown in Table.~\ref{Table1}, our system features an anisotropic charge conductivity and a symmetric spin conductivity tensor, while also hosting a finite second-order spin polarization. Moreover, the anisotropic spin splitting takes the form $J_0 k_x k_y + J_1 k_x k_y^3 + J_2 k_x^3 k_y$. This gives rise to the altermagnetic wrapping effect once $J_1\neq J_2$---a distinctive feature of our iron-chain altermagnet. It may also lead to additional spin degeneracies away from high-symmetry lines, e.g., at $k_y = \pm \sqrt{-J_0/J_1}$ (assuming $J_2 = 0$ and $\mathrm{sign}[J_0 J_1] < 0$).

Finally, we would also clarify the relation between our microscopic folding mechanism and flexomagnetism. A very recent work on the interplay between bending and altermagnetism demonstrated that curvature gradients in a thin altermagnetic film induce a local net magnetization~\cite{Yershov:prl25}. In our folded or bended chain with block-type order, the $[C_2||M_x]$ symmetry strictly links opposite magnetic moments and enforces exact compensation, yielding a pure altermagnetic state with zero net magnetization. Thus, while flexomagnetism offers a complementary geometric tuning knob, our mechanism is distinguished by the symmetry-protected absence of net magnetization.

In summary, we have demonstrated that altermagnetism can emerge in folded iron-chain systems exhibiting block-type magnetic order. Through a combination of symmetry analysis and first-principles calculations, we identified the critical role of lattice geometry, interchain coupling, and magnetic configuration in enabling non-relativistic spin splitting despite the absence of net magnetization. Our results reveal that the B1-type block state in Quasi-1D folded chains naturally breaks combined spacetime inversion symmetry, giving rise to altermagnetism. We further showed that this altermagnetic behavior is robust against chain rotation and sensitive to interchain spacing, highlighting the tunability of the effect. Quasi-1D K$_3$Fe$_2$Se$_4$ family materials could be an ideal platform to achieve block-type magnetic altermagnetism. Overall, this work establishes a unifying framework connecting real-space block magnetism and reciprocal-space altermagnetic phenomena and opens avenues for the design of novel quantum materials with symmetry-driven transport properties.

\noindent {\it Acknowledgements.--} 
\small {The work of L.-F. L., Y. Z., S. O., A. M. and E.D. were supported by the U.S. Department of Energy, Office of Science, Basic Energy Sciences, Materials Sciences and Engineering Division. We thank Bing Lv and Fei Gao for their inspiring discussions. This manuscript has been authored by UT-Battelle, LLC, under contract DE-AC05-00OR22725 with the US Department of Energy (DOE). The US government retains, and the publisher, by accepting the article for publication, acknowledges that the US government retains a nonexclusive, paid-up, irrevocable, worldwide license to publish or reproduce the published form of this manuscript, or allow others to do so, for US government purposes. DOE will provide public access to these results of federally sponsored research in accordance with the DOE Public Access Plan (https://www.energy.gov/doe-public-access-plan).}

{\it Data availability--}
The dataset of the main findings of this study is openly available in the Zenodo Repository (https://doi.org/10.5281/zenodo.17238044).

\noindent {\it Note added.--}
During the review of this manuscript, we became aware of a related work~\cite{Guo:prb25}, in which altermagnetism is controlled by tuning the inter-chain magnetic coupling between N\'eel chains through their separation. In contrast, we establish a strict 1D no-go theorem for altermagnetism and show that inter-chain hopping overcomes this constraint, driving PT-broken folded chains into a $d$-wave altermagnetic state [see End Matter].

\clearpage
\newpage
\begin{widetext}
\section{End Matter}
\begin{appendix}
\textit{Minimal model for the folded iron chain.---}
To capture the essential physics, we consider a minimal four-band tight-binding model for a folded iron chain with block-type magnetic order. The unit cell contains four magnetic sites: sites 1 and 4 carry spin-up moments, while sites 2 and 3 carry spin-down moments [see the SM~\cite{supp}]. In the basis $\Psi_{\mathbf{k}\sigma}^{\dagger}=(c_{1\mathbf{k}\sigma}^{\dagger}, c_{2\mathbf{k}\sigma}^{\dagger},c_{3\mathbf{k}\sigma}^{\dagger}, c_{4\mathbf{k}\sigma}^{\dagger})$, the Bloch Hamiltonian for spin sector $\sigma=\pm1$ takes the form
\begin{equation}
\mathcal{H}_{\sigma}(\mathbf{k})=
\begin{pmatrix}
-\sigma B & -t_{0}-t'e^{ik_{y}} & -t_{2}\bigl(1+e^{-ik_{x}}\bigr) & -t_{1}e^{-ik_{x}}-t_{3} \\[6pt]
-t_{0}-t'e^{-ik_{y}} & -\sigma B & -t_{1}-t_{3}e^{-ik_{x}} & -t_{2}\bigl(1+e^{-ik_{x}}\bigr) \\[6pt]
-t_{2}\bigl(1+e^{ik_{x}}\bigr) & -t_{1}-t_{3}e^{ik_{x}} & \sigma B & -t_{0}-t'e^{-ik_{y}} \\[6pt]
-t_{1}e^{ik_{x}}-t_{3} & -t_{2}\bigl(1+e^{ik_{x}}\bigr) & -t_{0}-t'e^{ik_{y}} & \sigma B
\end{pmatrix},
\label{eq:H_single_EM}
\end{equation}
where $B$ denotes the exchange splitting, $t_{0}$ the intracell nearest-neighbor hopping, $t_{1,2,3}$ the intrachain hoppings along the folded $x$ direction, and $t'$ the interchain (folding-induced vertical) hopping along $y$. The opposite spin sector follows from $\sigma\to-\sigma$.

To further illustrate the tunability of the altermagnetic spin splitting, we examine how the interchain distance $d$ affects the electronic structure. The interchain hopping decays exponentially with $d$, $t'(d)=0.75\,t_0\exp[-(d-d_0)/\lambda]$, where $d_0=1.0$ and $\lambda=1.0$ in lattice units. As shown in Fig.~\ref{fig:distance}(a), for $t'/t_0=0.75$, the spin-up and spin-down bands near $E\approx 0$ are split, opening a gap along the $\Gamma$--$M$ line. Figure~\ref{fig:distance}(b) shows that both $t'$ and the resulting spin-split gap decrease rapidly as $d$ increases. This behavior indicates that the spin splitting in the folded iron chain is primarily controlled by the interchain coupling: a larger interchain distance weakens $t'$, thereby reducing the spin-split gap. Our results thus suggest that the altermagnetic response can be effectively tuned by modifying the interchain spacing.

\begin{figure}[!htbp]
\centering
\includegraphics[width=0.8\textwidth]{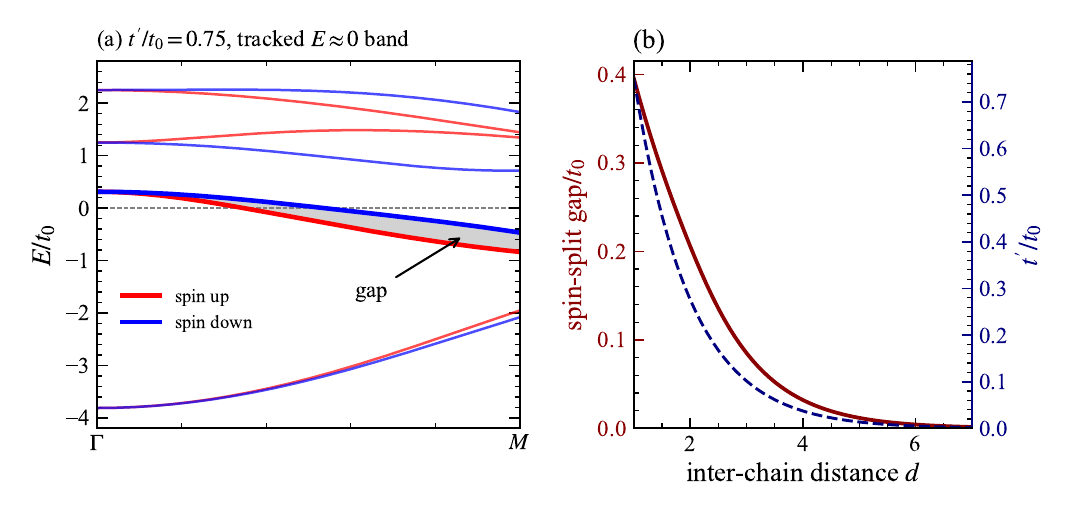}
\caption{
(a) Band structure of the folded iron chain along $\Gamma$--$M$ for $t'/t_0=0.75$, tracking the bands near $E\approx 0$. The red and blue lines denote spin-up and spin-down states, respectively. A spin-split gap opens near $E\approx 0$.
(b) Spin-split gap (red solid line, left axis) and interchain hopping $t'/t_0$ (blue dashed line, right axis) as functions of the interchain distance $d$. Both quantities decrease rapidly with increasing $d$.}
\label{fig:distance}
\end{figure}

In the above calculation, we have kept the interchain coupling FM. Consequently, a large interchain distance $d \gg 1$ destroys the altermagnetic spin splitting. This is the altermagnetic version of the ``no-go'' theorem for the 2D-to-1D crossover. It differs from Ref.~\cite{Guo:prb25}, where the vanishing of spin splitting with increasing $d$ originates from a transition of the interchain coupling from FM to AFM, thereby restoring the combined inversion and time-reversal symmetry [see Fig.~2(b-e) of Ref.~\cite{Guo:prb25}]. Thus, the two mechanisms are fundamentally different.
\end{appendix}
\end{widetext}


\begin{references}
\bibitem{Smejkal:prx1} L. Šmejkal, J. Sinova, and T. Jungwirth, Beyond conventional ferromagnetism and antiferromagnetism: a phase with nonrelativistic spin and crystal rotation symmetry. \href{https://doi.org/10.1103/PhysRevX.12.031042}{Phys. Rev. X \textbf{12}, 031042 (2022).}
\bibitem{Smejkal:prx2} L. Šmejkal, J. Sinova, and T. Jungwirth, Emerging research landscape of altermagnetism. \href{https://doi.org/10.1103/PhysRevX.12.040501}{Phys. Rev. X \textbf{12}, 040501 (2022).}
\bibitem{Mazin:prx}  I. Mazin, Editorial: altermagnetism — a new punch line of fundamental magnetism. \href{https://doi.org/10.1103/PhysRevX.12.040002} {Phys. Rev. X \textbf{12}, 040002 (2022).}
\bibitem{Song:nrm} Song, C., H. Bai, Z. Zhou, L. Han, H. Reichlova, J. H. Dil, , J. Liu, X. Chen, and  F. Pan, Altermagnets as a new class of functional materials. \href{https://doi.org/10.1038/s41578-025-00779-1} {Nat. Rev. Mater. \textbf{10}, 473–485 (2025).}
\bibitem{Naka:nc} M. Naka, S. Hayami, H. Kusunose, Y. Yanagi, Y. Motome, and H. Seo, Spin current generation in organic antiferromagnets. \href{https://doi.org/10.1038/s41467-019-12229-y} {Nat. Commun. \textbf{10}, 4305 (2019).}
\bibitem{Smejkal:sa} L. Šmejkal, R. González-Hernández, T. Jungwirth, and J. Sinova, Crystal time-reversal symmetry breaking and spontaneous Hall effect in collinear antiferromagnets. \href{https://doi.org/10.1126/sciadv.aaz8809} {Sci. Adv. \textbf{6}, eaaz8809 (2020).}
\bibitem{Ahn:prb} K.-H. Ahn, A. Hariki, K.-W. Lee, and J. Kuneš, Antiferromagnetism in RuO$_2$ as $d$-wave Pomeranchuk instability. \href{https://doi.org/10.1103/PhysRevB.99.184432} {Phys. Rev. B \textbf{99}, 184432 (2019).}
\bibitem{Yuan:prb} L. D. Yuan, Z. Wang, J. W. Luo, E. I. Rashba, A. Zunger,  Giant momentum-dependent spin splitting in centrosymmetric low-Z antiferromagnets. \href{https://doi.org/10.1103/PhysRevB.102.014422}{Phys. Rev. B \textbf{102}, 014422 (2020).}
\bibitem{Hayami:prb} S. Hayami, Y. Yanagi, and H. Kusunose, Bottom-up design of spin-split and reshaped electronic band structures in antiferromagnets without spin–orbit coupling: procedure on the basis of augmented multipoles. \href{https://doi.org/10.1103/PhysRevB.102.144441} {Phys. Rev. B \textbf{102}, 144441 (2020).}
\bibitem{Ma:nc} H.-Y. Ma, M. Hu, N. Li, J. Liu, W. Yao, J.F. Jia and J. Liu, Multifunctional antiferromagnetic materials with giant piezomagnetism and noncollinear spin current. \href{https://doi.org/10.1038/s41467-021-23127-7}{Nat. Commun.  \textbf{12}, 2846 (2021).}
\bibitem{Mazin:pnas} I. I. Mazin, K. Koepernik, M. D. Johannes, R. González-Hernández and L. Šmejkal, Prediction of unconventional magnetism in doped FeSb$_2$. \href{https://doi.org/10.1073/pnas.2108924118} {Proc. Natl Acad. Sci. USA  \textbf{118}, e2108924118 (2021).}
\bibitem{Gao:nsr2025} Z.-F. Gao, S. Qu, B. Zeng, Y. Liu, J.-R. Wen, H. Sun, P.-J. Guo, and Z.-Y. Lu, AI-accelerated discovery of altermagnetic materials. \href{https://doi.org/10.1093/nsr/nwaf066}{Natl. Sci. Rev. \textbf{12}, nwaf066 (2025).}
\bibitem{Liu:prx} P. Liu, J. Li, J. Han, X. Wan and Q. Liu,  Spin-group symmetry in magnetic materials with negligible spin–orbit coupling. \href{https://doi.org/10.1103/PhysRevX.12.021016}{Phys. Rev. X \textbf{12}, 021016 (2022).}
\bibitem{Antonenko:prl2025} D. S. Antonenko, R. M. Fernandes, and J. W. F. Venderbos, Mirror Chern Bands and Weyl Nodal Loops in Altermagnets. \href{https://doi.org/10.1103/PhysRevLett.134.096703}{Phys. Rev. Lett. \textbf{134}, 096703 (2025).}
\bibitem{Fernandes:prb2024} R. M. Fernandes, V. S. De Carvalho, T. Birol, and R. G. Pereira, Topological transition from nodal to nodeless Zeeman splitting in altermagnets. \href{https://doi.org/10.1103/PhysRevB.109.024404}{Phys. Rev. B \textbf{109}, 024404 (2024).} 
\bibitem{Zhou:prb2025} Y. Zhou, C.-A. Ji, S. Dong, and X.-Z. Lu, Out-of-plane ferroelectricity, magnetoelectric coupling, and persistent spin texture in two-dimensional multiferroics. \href{https://doi.org/10.1103/rcg6-3gxl}{Phys. Rev. B \textbf{112}, 054412 (2025).}
\bibitem{Li:pnas2024} H. Li, G. Wang, N. Ding, Q. Ren, G. Zhao, W. Lin, J. Yang, W. Yan, Q. Li, R. Yang, S. Yuan, J. D. Denlinger, Z. Wang, X. Zhang, L. A. Wray, S. Dong, D. Qian, and L. Miao, Spectroscopic evidence of spin-state excitation in d-electron correlated semiconductor FeSb2, \href{https://doi.org/10.1073/pnas.2321193121} {Proc. Natl. Acad. Sci. U.S.A. \textbf{121}, e2321193121 (2024).}
\bibitem{Jungwirth:arXiv2024} T. Jungwirth, R.M. Fernandes, J. Sinova, and L. Smejkal, Altermagnetism: An unconventional spin-ordered phase of matter, \href{https://doi.org/10.1016/j.newton.2025.100162} {Newton \textbf{1}, 100162 (2025)}
\bibitem{Fender:jacs2025} S. S. Fender, O. Gonzalez, and D. K. Bediako, Altermagnetism: A chemical perspective, \href{https://doi.org/10.1021/jacs.4c14503} {J. Am. Chem. Soc. \textbf{147}, 2257 (2025).}
\bibitem{Naka:prb2021} M. Naka, Y. Motome, and H. Seo, Perovskite as a spin current generator. \href{https://doi.org/10.1103/PhysRevB.103.125114}{Phys. Rev. B \textbf{103}, 125114 (2021).}
\bibitem{Shao:NatComm} D.-F. Shao, S.-H. Zhang, M. Li, C.-B. Eom, E. Y. Tsymbal, Spin-neutral currents for spintronics. \href{https://doi.org/10.1038/s41467-021-26915-3}{Nat. Commun. \textbf{12}, 7061 (2021).}
\bibitem{Zhang:prl2024Predictable} R. W. Zhang, C. Cui, R. Li, J. Duan, L. Li, Z.-M. Yu, and Y. Yao, Predictable Gate-Field Control of Spin in Altermagnets with Spin-Layer Coupling. \href{https://doi.org/10.1103/PhysRevLett.133.056401}{Phys. Rev. Lett. \textbf{133}, 056401 (2024).}
\bibitem{Bhowal:prx} S. Bhowal and N. A. Spaldin, Ferroically ordered magnetic octupoles in d-wave altermagnets. \href{https://doi.org/10.1103/PhysRevX.14.011019}{Phys. Rev. X \textbf{14}, 011019 (2024).}
\bibitem{Leeb:prl2024} V. Leeb, A. Mook, L. Šmejkal, and J. Knolle, Spontaneous Formation of Altermagnetism from Orbital Ordering. \href{https://doi.org/10.1103/PhysRevLett.132.236701}{Phys. Rev. Lett. \textbf{132}, 236701 (2024).}
\bibitem{Duan:prl2025} X. Duan, J. Zhang, Z. Zhu, Y. Liu, Z. Zhang, I. Žutić, and T. Zhou, Antiferroelectric Altermagnets: Antiferroelectricity Alters Magnets. \href{https://doi.org/10.1103/PhysRevLett.134.106801}{Phys. Rev. Lett. \textbf{134}, 106801 (2025).} 
\bibitem{Gu:prl2025} M. Gu, Y. Liu, H. Zhu, K. Yananose, X. Chen, Y. Hu, A. Stroppa, and Q. Liu, Ferroelectric Switchable Altermagnetism. \href{https://doi.org/10.1103/PhysRevLett.134.106802}{Phys. Rev. Lett. \textbf{134}, 106802 (2025).}
\bibitem{Zhu:NanoLett2025} Z. Zhu, X. Duan, J. Zhang, B. Hao, I. Žutić, and T. Zhou, Two-Dimensional Ferroelectric Altermagnets: From Model to Material Realization. \href{https://doi.org/10.1021/acs.nanolett.5c02121}{Nano Lett. \textbf{25}, 9456-9462 (2025).}
\bibitem{Maier:prb2023Weak} T. A. Maier and S. Okamoto, Weak-coupling theory of neutron scattering as a probe of altermagnetism. \href{https://doi.org/10.1103/PhysRevB.108.L100402}{Phys. Rev. B \textbf{108}, L100402 (2023).}
\bibitem{Bose:prb2024} A. Bose, S. Vadnais, and A. Paramekanti, Altermagnetism and superconductivity in a multiorbital \(t\!-\!J\) model. \href{https://doi.org/10.1103/PhysRevB.110.205120}{Phys. Rev. B \textbf{110}, 205120 (2024).}
\bibitem{Kaushal:arXiv} N. Kaushal and M. Franz, Altermagnetism in modified Lieb lattice Hubbard model. \href{https://doi.org/10.1103/s31h-hk2v}{Phys. Rev. Lett. \textbf{135}, 156502 (2025).}
\bibitem{Wu:PRL2004} C. Wu, and S.-C. Zhang, Dynamic Generation of Spin-Orbit Coupling. \href{https://doi.org/10.1103/PhysRevLett.93.036403}{Phys. Rev. Lett. \textbf{93}, 036403 (2004).}
\bibitem{Wu:PRB2007} C. Wu, K. Sun, E. Fradkin, and S.-C. Zhang, Fermi liquid instabilities in the spin channel. \href{https://doi.org/10.1103/PhysRevB.75.115103}{Phys. Rev. B \textbf{75}, 115103  (2007).}
\bibitem{Jungwirth:nature} T. Jungwirth, J. Sinova, R. M. Fernandes, Q. Liu, H. Watanabe, S. Murakami, S. Nakatsuji, and L. \v{S}mejkal, Symmetry, microscopy and spectroscopy signatures of altermagnetism. \href{ https://doi.org/10.1038/s41586-025-09883-2}{Nautre \textbf{649}, 837 (2026).}
\bibitem{Xiao:prx} Z. Xiao, J. Zhao, Y. Li, R. Shindou, and Z.-D. Song, Spin Space Groups: Full Classification and Applications. \href{https://doi.org/10.1103/PhysRevX.14.031037} {Phys. Rev. X \textbf{14}, 031037 (2024).}
\bibitem{Chen:prx} X. Chen, J. Ren, Y. Zhu, Y. Yu, A. Zhang, P. Liu, J. Li, Y. Liu, C. Li, and Q. Liu, Enumeration and Representation Theory of Spin Space Groups. \href{https://doi.org/10.1103/PhysRevX.14.031038} {Phys. Rev. X \textbf{14}, 031038 (2024).}
\bibitem{Jiang:prx} Y. Jiang, Z. Song, T. Zhu, Z. Fang, H. Weng, Z.-X. Liu, J. Yang, and C. Fang, Enumeration of Spin-Space Groups: Toward a Complete Description of Symmetries of Magnetic Orders. \href{https://doi.org/10.1103/PhysRevX.14.031039}{Phys. Rev. X \textbf{14}, 031039 (2024).}
\bibitem{Feng:ner} Z. Feng, X. Zhou, L. Šmejkal, L. Wu, Z. Zhu, H. Guo, R. González-Hernández, X. Wang, H. Yan, P. Qin, X. Zhang, H. Wu, H. Chen, Z. Meng, L. Liu, Z. Xia, J. Sinova, T. Jungwirth, and Z. Liu, An anomalous Hall effect in altermagnetic ruthenium dioxide. \href{https://doi.org/10.1038/s41928-022-00866-z}{Nat. Electron. \textbf{5}, 735 (2022).}
\bibitem{Han:sciadv2024b} L. Han, Z. Liu, Y. Chen, H. Song, Y. Xu, X. Zhang, and S. Wang, Electrical 180° switching of Néel vector in spin-splitting antiferromagnet. \href{https://doi.org/10.1126/sciadv.adn0479}{Sci. Adv. \textbf{10}, eadn0479 (2024).}
\bibitem{Han:PRApplied2025} L. Han, X. Z. Fu, W. Q. He, J. K. Dai, Y. X. Zhu, W. F. Yang, Y. L. Chen, J. C. Zhang, W. X. Zhu, H. Bai, C. Chen, D. Z. Hou, C. H. Wan, X. F. Han, C. Song, J. W. Liu, and F. Pan, Nonvolatile anomalous Nernst effect in Mn$_5$Si$_3$ with a collinear N\'eel vector, \href{https://doi.org/10.1103/PhysRevApplied.23.044066} {Phys. Rev. Applied \textbf{23}, 044066 (2025).}
\bibitem{Krempasky:nat2024} J. Krempaský, L. Šmejkal, S. W. D’Souza, M. Hajlaoui, G. Springholz, K. Uhlířová, F. Alarab, P. C. Constantinou, V. Strocov, D. Usanov, W. R. Pudelko, R. González-Hernández, A. Birk Hellenes, Z. Jansa, H. Reichlová, Z. Šobáň, R. D. Gonzalez Betancourt, P. Wadley, J. Sinova, D. Kriegner, J. Minár, J. H. Dil and T. Jungwirth, Altermagnetic lifting of Kramers spin degeneracy. \href{https://doi.org/10.1038/s41586-023-06907-7}{Nature \textbf{626}, 517 (2024).}
\bibitem{Fedchenko:sciadv2024} O. Fedchenko, J. Minár, A. Akashdeep, S. W. D’Souza, D. Vasilyev, O. Tkach, L. Odenbreit, Q. Nguyen, D. Kutnyakhov, N. Wind, L. Wenthaus, M. Scholz, K. Rossnagel, M. Hoesch, M. Aeschlimann, B. Stadtmüller, M. Kläui, G. Schönhense, T. Jungwirth, A. B. Hellenes, G. Jakob, L. Šmejkal, J. Sinova, and H.-J. Elmers, Observation of time-reversal symmetry breaking in the band structure of altermagnetic RuO$_2$. \href{https://doi.org/10.1126/sciadv.adj4883}{Sci. Adv. \textbf{10}, eadj4883 (2024).}
\bibitem{Lee:prl2024} S. Lee, S. Lee, S. Jung, J. Jung, D. Kim, Y. Lee, B. Seok, J. Kim, B. G. Park, L. Šmejkal, C.-J. Kang, and C. Kim, Broken Kramers Degeneracy in Altermagnetic MnTe. \href{https://doi.org/10.1103/PhysRevLett.132.036702}{Phys. Rev. Lett. \textbf{132}, 036702 (2024).}
\bibitem{Ding:prl2024} J. Ding, Z. Jiang, X. Chen, Z. Tao, Z. Liu, T. Li, J. Liu, J. Sun, J. Cheng, J. Liu, Y. Yang, R. Zhang, L. Deng, W. Jing, Y. Huang, Y. Shi, M. Ye, S. Qiao, Y. Wang, Y. Guo, D. Feng, and D. Shen, Large band splitting in g‑wave altermagnet CrSb. \href{https://doi.org/10.1103/PhysRevLett.133.206401}{Phys. Rev. Lett. \textbf{133}, 206401 (2024).}
\bibitem{Zeng:adv2024} M. Zeng, M.-Y. Zhu, Y.-P. Zhu, X.-R. Liu, X.-M. Ma, Y.-J. Hao, P. Liu, G. Qu, Y. Yang, Z. Jiang, K. Yamagami, M. Arita, X. Zhang, T.-H. Shao, Y. Dai, K. Shimada, Z. Liu, M. Ye, Y. Huang, Q. Liu, and C. Liu, Observation of spin splitting in room‑temperature metallic antiferromagnet CrSb. \href{https://doi.org/10.1002/advs.202406529}{Adv. Sci. \textbf{11}, 2406529 (2024).}
\bibitem{Yang:natcomm2025}G. Yang, Z. Li, S. Yang, J. Li, H. Zheng, W. Zhu, Z. Pan, Y. Xu, S. Cao, W. Zhao, A. Jana, J. Zhang, M. Ye, Y. Song, L.-H. Hu, L. Yang, J. Fujii, I. Vobornik, M. Shi, H. Yuan, Y. Zhang, Y. Xu, and Y. Liu, Three-dimensional mapping of the altermagnetic spin splitting in CrSb. \href{https://doi.org/10.1038/s41467-025-56647-7}{Nat. Commun. \textbf{16}, 1442 (2025).}
\bibitem{Reimers:natcomm2024} S. Reimers, L. Odenbreit, L. Šmejkal, V. N. Strocov, P. Constantinou, A. B. Hellenes, R. Jaeschke Ubiergo, W. H. Campos, V. K. Bharadwaj, A. Chakraborty, T. Denneulin, W. Shi, R. E. Dunin-Borkowski, S. Das, M. Kläui, J. Sinova, and M. Jourdan, Direct observation of altermagnetic band splitting in CrSb thin films. \href{https://doi.org/10.1038/s41467-024-46476-5}{Nat. Commun. \textbf{15}, 2116 (2024).}
\bibitem{Osumi:prb2024} T. Osumi, S. Souma, T. Aoyama, K. Yamauchi, A. Honma, K. Nakayama, T. Takahashi, K. Ohgushi, and T. Sato, Observation of a giant band splitting in altermagnetic MnTe. \href{https://doi.org/10.1103/PhysRevB.109.115102}{Phys. Rev. B \textbf{109}, 115102 (2024).}
\bibitem{Liu:prl2024chiral} Z. Liu, M. Ozeki, S. Asai, S. Itoh, and T. Masuda, Chiral split magnon in altermagnetic MnTe, \href{https://doi.org/10.1103/PhysRevLett.133.156702} {Phys. Rev. Lett. \textbf{133}, 156702 (2024).}
\bibitem{Liu:prl2024} J. Liu, J. Zhan, T. Li, J. Liu, S. Cheng, Y. Shi, L. Deng, M. Zhang, C. Li, J. Ding, Q. Jiang, M. Ye, Z. Liu, Z. Jiang, S. Wang, Q. Li, Y. Xie, Y. Wang, S. Qiao, J. Wen, Y. Sun, and D. Shen, Absence of altermagnetic spin splitting in rutile RuO$_2$. \href{https://doi.org/10.1103/PhysRevLett.133.176401}{Phys. Rev. Lett. \textbf{133}, 176401 (2024).}
\bibitem{Jiang:nphys2025} B. Jiang, M. Hu, J. Bai, Z. Song, C. Mu, G. Qu, W. Li, W. Zhu, H. Pi, Z. Wei, Y.-J. Sun, Y. Huang, X. Zheng, Y. Peng, L. He, S. Li, J. Luo, Z. Li, G. Chen, H. Weng, and T. Qian, A metallic room‑temperature d‑wave altermagnet. \href{https://doi.org/10.1038/s41567-025-02822-y}{Nat. Phys. \textbf{21}, 754–759 (2025).}
\bibitem{Zhang:arxiv2024} F. Zhang, X. Cheng, Z. Yin, C. Liu, L. Deng, Y. Qiao, Z. Shi, S. Zhang, J. Lin, Z. Liu, M. Ye, Y. Huang, X. Meng, C. Zhang, T. Okuda, K. Shimada, S. Cui, Y. Zhao, G.-H. Cao, S. Qiao, J. Liu, and C. Chen, Crystal-symmetry-paired spin–valley locking in a layered room-temperature metallic altermagnet candidate. \href{https://doi.org/10.1038/s41567-025-02864-2}{Nat. Phys. \textbf{21}, 760 (2025).}
\bibitem{Mourigal:prl2015} M. Mourigal, S. Wu, M. B. Stone, J. R. Neilson, J. M. Caron, T. M. McQueen, and C. L. Broholm, Block magnetic excitations in the orbitally-selective Mott insulator BaFe$_2$Se$_3$. \href{https://doi.org/10.1103/PhysRevLett.115.047401}{Phys. Rev. Lett. \textbf{115}, 047401 (2015).}
\bibitem{Ye:prl2011} F. Ye, S. Chi, W. Bao, X. F. Wang, J. J. Ying, X. H. Chen, H. D. Wang, C. H. Dong, and M. Fang, Common crystalline and magnetic structure of superconducting A$_2$Fe$_4$Se$_5$ (A = K, Rb, Cs, Tl) single crystals measured using neutron diffraction.
\href{https://doi.org/10.1103/PhysRevLett.107.137003}{Phys. Rev. Lett. \textbf{107}, 137003 (2011).}
\bibitem{Dagotto:rmp2013} E. Dagotto, The unexpected properties of alkali metal iron selenide superconductors. \href{https://doi.org/10.1103/RevModPhys.85.849}{Rev. Mod. Phys. \textbf{85}, 849–867 (2013).}
\bibitem{Lin:prl2021} L. F. Lin, Y. Zhang, G. Alvarez, A. Moreo, and E. Dagotto, Origin of Insulating Ferromagnetism in Iron Oxychalcogenide Ce$_2$O$_2$FeSe$_2$.\href{https://doi.org/10.1103/PhysRevLett.127.077204}{Phys. Rev. Lett. \textbf{127}, 077204 (2021).}
\bibitem{Lin:prb2022} L.-F. Lin, Y. Zhang, G. Alvarez, A. Moreo, and E. Dagotto, Prediction of orbital-selective Mott phases and block magnetic states in the quasi-one-dimensional iron chain Ce$_2$O$_2$FeSe$_2$ under hole and electron doping. \href{https://doi.org/10.1103/PhysRevB.105.075119}{Phys. Rev. B \textbf{105}, 075119 (2022).}
\bibitem{Lin:cp2023} L. F. Lin, Y. Zhang, G. Alvarez, M. A. McGuire, A. F. May, A. Moreo, and E. Dagotto, Stability of the interorbital-hopping mechanism for ferromagnetism in multi-orbital Hubbard models. \href{https://doi.org/10.1038/s42005-023-01314-w}{Commun. Phys. \textbf{6}, 199 (2023).}
\bibitem{Fu:prl2007} L. Fu, C. L. Kane, and E. J. Mele, Topological Insulators in Three Dimensions. \href{https://doi.org/10.1103/PhysRevLett.98.106803}{Phys. Rev. Lett. \textbf{98}, 106803 (2007).}
\bibitem{Zhuang:arxiv2025} Z.-Y. Zhuang, D. Zhu, D. Liu, Z. Wu, Z. Yan, Odd-Parity Altermagnetism Originated from Orbital Orders. \href{https://doi.org/10.48550/arXiv.2508.18361}{arXiv 2508.18361 (2025).}
\bibitem{supp}See Supplemental Material for method details, more DFT and tight-binding results, and transport calculations.
\bibitem{Momma:vesta} K. Momma and F. Izumi, \href{https://doi.org/10.1107/S0021889811038970}{J. Appl. Crystallogr. \textbf{44}, 1272 (2011).}

\bibitem{Kresse:prb1993} G. Kresse and J. Hafner, Ab initio molecular dynamics for liquid metals. \href{https://doi.org/10.1103/PhysRevB.47.558}{Phys. Rev. B \textbf{47}, 11169–11186 (1993).}
\bibitem{Blochl:Prb} P. E. Bl\"{o}chl, Projector augmented-wave method \href{https://doi.org/10.1103/PhysRevB.50.17953}{Phys. Rev. B \textbf{50}, 17953 (1994).}
\bibitem{Kresse:prb1996} G. Kresse and J. Furthmüller, Efficient iterative schemes for $ab~initio$ total-energy calculations using a plane-wave basis set. \href{https://doi.org/10.1103/PhysRevB.54.11169}{Phys. Rev. B \textbf{54}, 11169–11186 (1996).}
\bibitem{Kresse:cms1996} G. Kresse and J. Furthmüller, Efficiency of $ab~initio$ total energy calculations for metals and semiconductors using a plane-wave basis set. \href{https://doi.org/10.1016/0927-0256(96)00008-0}{Comput. Mater. Sci. \textbf{6}, 15–50 (1996).}
\bibitem{Perdew:prl1996} J. P. Perdew, K. Burke, and M. Ernzerhof, Generalized Gradient Approximation Made Simple. \href{https://doi.org/10.1103/PhysRevLett.77.3865} {Phys. Rev. Lett. \textbf{77}, 3865 (1996).}
\bibitem{Dudarev:prb1998} S. L. Dudarev, G. A. Botton, S. Y. Savrasov, C. J. Humphreys, and A. P. Sutton, Electron-energy-loss spectra and the structural stability of nickel oxide: An LSDA+U study. \href{https://doi.org/10.1103/PhysRevB.57.1505} {Phys. Rev. B \textbf{57}, 1505 (1998).}
\bibitem{McCabe:prb} E. E. McCabe, C. Stock, J. L. Bettis, Jr., M.-H. Whangbo, and J. S. O. Evans, Magnetism of the Fe$^{2+}$ and Ce$^{3+}$ sublattices in Ce$_2$O$_2$FeSe$_2$: A combined neutron powder diffraction, inelastic neutron scattering, and density functional study. \href{https://doi.org/10.1103/PhysRevB.90.235115}{Phys. Rev. B \textbf{90}, 235115 (2014).}
\bibitem{Zhang:prb} Y. Zhang, P. Laurell, G. Alvarez, A. Moreo, T. A. Maier, L.-F. Lin, and E. Dagotto, Intertwined charge, spin, and orbital degrees of freedom under electronic correlations in the one-dimensional Fe$^{3+}$ chalcogenide chain. \href{https://doi.org/10.1103/th25-rvx7}{Phys. Rev. B \textbf{113}, 245120 (2026).}
\bibitem{Gao:prl} Fei Gao, Nikhil Dhale, Ling-Fang Lin, Keith M. Taddei, Yang Zhang, Clarina Dela Cruz, Elbio Dagotto, and Bing Lv, Block-type antiferromagnetism in quasi-one-dimensional K$_3$Fe$_2$Se$_4$. \href{https://doi.org/10.1103/g8zx-gc2t}{Phys. Rev. Lett.  \textbf{136}, 256706 (2026).}
\bibitem{Li:prb25} Z. Li, X. Ma, S. Wu, H.-Q. Yuan, J. Dai, and C. Cao, Layered ternary iron-selenides under pressure as candidates for coexisting altermagnetism and superconductivity. \href{https://doi.org/10.1103/tst1-924w}{Phys. Rev. B \textbf{112}, 214456  (2025).}
\bibitem{Zhu:prl25} X. Zhu, X. Huo, S. Feng, S.-B. Zhang, S. A. Yang, and H. Guo, Design of Altermagnetic Models from Spin Clusters. \href{https://doi.org/10.1103/PhysRevLett.134.166701}{Phys. Rev. Lett. \textbf{134}, 166701 (2025).}
\bibitem{Stueble:zaac2017} P. Stüble and C. Röhr, Cs[FeSe$_2$], Cs$_3$[FeSe$_2$]$_2$, and Cs$_7$[Fe$_4$Se$_8$]: missing links of known chalcogenido-ferrate series. \href{https://doi.org/10.1002/zaac.201700263}{Z. Anorg. Allg. Chem. \textbf{643}, 1462–1473 (2017).}
\bibitem{Harrison:ccr} M. R. Harrison and M. G. Francesconi, Mixed-metal one-dimensional sulfides—A class of materials with differences and similarities to oxides. \href{https://doi.org/10.1016/j.ccr.2010.10.008}{Coord. Chem. Rev. \textbf{255}, 451–458 (2011).}
\bibitem{Bronger:jalloy} W. Bronger, U. Ruschewitz, and P. Müller, New ternary iron sulphides $A_3$Fe$_2$S$_4$ ($A$ = K, Rb, Cs): syntheses and crystal structures. \href{https://doi.org/10.1016/0925-8388(94)01344-6}{J. Alloys Compd. \textbf{218}, 22 (1995).}
\bibitem{Klepp:kryst2000} K. O. Klepp, S. Pantschov, and H. Boller, Crystal structure of mixed-valent trirubidium tetraselenidodiferrate, Rb$_3$Fe$_2$Se$_4$. \href{https://doi.org/10.1515/ncrs-2000-0105}{Z. Kristallogr. – New Cryst. Struct. \textbf{215}, 5–6 (2000).}
\bibitem{Bronger:zaac1999} W. Bronger, H. S. Genin, and P. Müller, Crystal structure and polymorphism of K$_3$Fe$_2$Se$_4$. \href{https://doi.org/10.1002/(SICI)1521-3749(199902)625:2<274::AID-ZAAC274>3.0.CO;2-2}{Z. Anorg. Allg. Chem. \textbf{625}, 274–278 (1999).}
\bibitem{Ensling:hfi} J. Ensling, P. Gütlich, and H. Spiering, Mössbauer and magnetic studies of mixed-valence linear chain compounds: Na$_3$Fe$_2$S$_4$ and Na$_3$Fe$_2$Se$_4$ \href{https://doi.org/10.1007/BF02061519}{Hyperfine Interactions \textbf{28}, 599 (1986).}
\bibitem{Welz:jpcs1989} D. Welz, P. Deppe, W. Schäfer, H. Sabrowsky, and M. Rosenberg, Magnetism of iron–sulfur tetrahedral frameworks in compounds with thallium I: chain structures. \href{https://doi.org/10.1016/0022-3697(89)90492-7}{J. Phys. Chem. Solids \textbf{50}, 297–308 (1989).}
\bibitem{Ghazanfari:rscadv2022} M. R. Ghazanfari, A. Santhosh, J. C. Vrijmoed, K. Siemensmeyer, B. Peters, S. Dehnen, P. Jerabek, and G. Thiele, Large-scale synthesis of mixed valence K$_3$Fe$_2$S$_4$ with high dielectric and ferrimagnetic characteristics. \href{https://doi.org/10.1039/D2RA05200H}{RSC Adv. \textbf{12}, 30514-30521 (2022).}
\bibitem{Berlijn:prl} T. Berlijn, P. C. Snijders3, O. Delaire3, H.-D. Zhou, T. A. Maier, H.-B. Cao, S.-X. Chi, M. Matsuda, Y. Wang, M. R. Koehler, P. R. C. Kent, and H. H. Weitering, Itinerant Antiferromagnetism in RuO$_2$. \href{https://doi.org/10.1103/PhysRevLett.118.077201}{Phys. Rev. Lett. \textbf{118}, 077201 (2017).}
\bibitem{Zhu:prl22} Z. H. Zhu, J. Strempfer, R. R. Rao, C. A. Occhialini, J. Pelliciari, Y. Choi, T. Kawaguchi, H. You, J. F. Mitchell, Y. Shao-Horn, and R. Comin, Anomalous Antiferromagnetism in Metallic RuO$_2$ Determined by Resonant X-ray Scattering. \href{https://doi.org/10.1103/PhysRevLett.122.017202}{Phys. Rev. Lett. \textbf{122}, 017202 (2019).}
\bibitem{Hiraishi:prl24} M. Hiraishi, H. Okabe, A. Koda, R. Kadono, T. Muroi, D. Hirai, and Z. Hiroi, Nonmagnetic Ground State in RuO$_2$ Revealed by Muon Spin Rotation. \href{https://doi.org/10.1103/PhysRevLett.132.166702}{Phys. Rev. Lett. \textbf{132}, 166702 (2024).}
\bibitem{Wang:arxiv2025-V} Z. Wang, S. Yu, X. Cheng, X. Xiao, W. Ma, F. Quan, H. Song, K. Zhang, Y. Zhang, Y. Ma, W. Liu, P. Yadav, X. Shi, Z. Wang, Q. Niu, Y. Gao, B. Xiang, J. Liu, Z. Wang, and X. Chen, Atomic-scale spin sensing of a 2D -wave altermagnet via helical tunneling. \href{https://doi.org/10.48550/arXiv.2512.23290}{arXiv, 2512.23290 (2025).}
\bibitem{Fu:arxiv2025-V} D. Fu, L. Yang, Y. Shen, K. Xiao, Y. Wang, W. Jiang, Z. Wang, Y. Yao, Q.-K. Xue, and W. Li, Atomic-scale visualization of $d$-wave altermagnetism. \href{https://doi.org/10.48550/arXiv.2512.24114}{arXiv, 2512.24114 (2025).}
\bibitem{Yang:arxiv2025-V} G. Yang, C. Li, C. Wang, X. Zhao, Y. Wan, H. Gui, G. Zeng, S. Cao, C. Hu, D. Chen, Y. Liu, Y. Song, F. Liu, L.-H. Hu, L. Jiao, and H. Yuan, Visualizing spin-polarization of an altermagnet KV$_2$Se$_2$O via spin-selective tunneling. \href{https://doi.org/10.48550/arXiv.2603.21969}{arXiv, 2603.21969 (2026).}
\bibitem{Mu:arxiv2025-V} D. Mu, B. Jiang, Q. Duan, Z. Xu, X. Cheng, Y. Xiao, X. Han, X. Liang, Z. Luo, R. L. Kong, Q. Wang, J. Liu, J. Zhong, R. Zhong, Q. Gu, B. Lv, and H. Ding, Direct observation of quadruple spin-texture locking in a 2D $d$-wave altermagnet. \href{https://doi.org/10.48550/arXiv.2604.18337}{arXiv, 2604.18337 (2026).}
\bibitem{Yershov:prl25} K. V. Yershov, O. Gomonay, J. Sinova, J. van den Brink, and V. P. Kravchuk, Curvature-Induced Magnetization of Altermagnetic Films. \href{https://doi.org/10.1103/PhysRevLett.134.116701}{Phys. Rev. Lett. \textbf{134}, 116701 (2025).}
\bibitem{Sun:prb2025} Y. Sun, Y. Huang, J. Cheng, S. Zhang, Z. Li, H. Luo, X. Ma, W. Yang, J. Yang, D. Chen, K. Sun, M. Gutmann, S. C. Capelli, F. Shen, J. Hao, L. He, G. Chen, and S. Li, Antiferromagnetic structure of KV$_2$Se$_2$O: A neutron diffraction study, \href{https://doi.org/10.1103/27n8-5q4l}{Phys. Rev. B \textbf{112}, 184416 (2025).}
\bibitem{Yang:arxiv2025} G. Yang, R. Chen, C. Liu, J. Li, Z. Pan, T. Hua, P. Xiong, L. Deng, N. Zheng, Y. Tang, H. Zheng, W. Zhu, Y. Xu, X. Zheng, X. Ma, X. Wang, S. Cui, Z. Sun, Z. Liu, M. Ye, C. Cao, M. Shi, L. Hu, Q. Liu, S. Qiao, G. Cao, Y. Song, and Y. Liu, Observation of hidden altermagnetism in Cs$_{1-\delta}$V$_2$Te$_2$O, \href{https://doi.org/10.48550/arXiv.2512.00972}{arXiv:2512.00972v2 (2026).}
\bibitem{Li:arxiv2025} C. Li, J.-X. Hou, S.-L. Zhu, H. Zheng, Y. Song, Y. Liu, S.-B. Zhang, L.-H. Hu, Altermagnetic Even-Odd Effects in CsV$_2$Te$_2$O Josephson Junctions. \href{https://doi.org/10.48550/arXiv.2602.14485}{arXiv:2602.14485, (2026).}
\bibitem{Guo:fp} S.-D. Guo, Hidden altermagnetism. \href{https://doi.org/10.15302/frontphys.2026.025201}{Front. Phys. \textbf{21}, 025201 (2026).}
\bibitem{Matsuda:prl} J. Matsuda, H. Watanabe, and R. Arita, Multiferroic Collinear Antiferromagnets with Hidden Altermagnetic Spin Splitting. \href{https://doi.org/10.1103/vgcs-bn8g}{Phys. Rev. Lett.  \textbf{134}, 226703 (2025).}
\bibitem{Meier:prl} Q. N. Meier, A. Carta, C. Ederer, and A. Cano, Net and Compensated Altermagnetism from Staggered Orbital Order: Layer-Dependent Spin Splitting in Sr$_{n+1}$Cr$_n$O$_{3n+1}$. \href{https://doi.org/10.1103/vgcs-bn8g}{Phys. Rev. Lett. \textbf{136}, 116705 (2026).}
\bibitem{Guo:prb25} D. Guo, C. Zong, W. Zhang, C. Wang, J. Liu, and W. Ji, Tunable altermagnetism via interchain engineering in parallel-assembled atomic chains. \href{https://doi.org/10.1103/zhds-vnyt}{Phys. Rev. B \textbf{112}, L041404 (2025).}
\end{references}
\end{document}